\documentclass[]{aa}
\usepackage{graphicx}
\usepackage{txfonts}
\usepackage{booktabs}
\usepackage{multirow}

\usepackage[table,xcdraw]{xcolor}
\usepackage{xspace}

\begin{document}

   \title{The current cratering rate on the regular satellites of Jupiter, Saturn, and Uranus}

   \author{R. Brasser\inst{1}\fnmsep\inst{2}\thanks{rbrasser@konkoly.hu},
          E. W. Wong\inst{3}\fnmsep\inst{4}
          \and
          S. C. Werner\inst{2}}
   \institute{Konkoly Observatory, HUN-REN CSFK, MTA Centre of Excellence; Konkoly Thege Miklos St. 15-17, H-1121 Budapest, Hungary  
    \and 
   Centre for Planetary Habitability (PHAB), Department of Geosciences, University of Oslo; Sem Saelands Vei 2A, N-0371 Oslo, Norway
        \and
        Earth Life Science Institute, Tokyo Institute of Technology, Meguro-ku, Tokyo, 152-8550, Japan
        \and
        Observatoire de l’Université de Genève, 51 chemin des Maillettes, 1290 Versoix, Switzerland}
\authorrunning{Brasser, Wong and Werner}
   \date{}

\abstract
{The impact or cratering rates onto the regular satellites of the giant planets is subject to great uncertainties. A better knowledge of the impact rate is useful to establish the surface ages of young terrains on these satellites, such as on Europa, Enceladus, Tethys and Dione.}
{We aim to compute the impact rates for objects with a diameter of 1 km onto the regular satellites of Jupiter, Saturn and Uranus using our latest dynamical simulations of the evolution of outer solar system coupled with the best estimates of the current population of objects beyond Neptune and their size-frequency distribution.}
{We use the outcome of the last 3.5~Gyr of evolution of the outer solar system from our database of simulations and combine this with observational constraints of the population beyond Neptune to compute the flux of objects entering the Centaur region, with uncertainties. The initial conditions resemble the current population rather than a near-circular, near-planar disc usually assumed just before the onset of giant planet migration. We obtain a better estimate of the impact probability of a Centaur with the satellites from enacting simulations of planetesimals flying past the satellites on hyperbolic orbits, which agree with literature precedents.} 
{We find that our impact rate of objects greater than 1 km in diameter with Jupiter is 0.0012~yr$^{-1}$, which is a factor of 3--6 lower than previous estimates of 0.0044~yr$^{-1}$ \citep{Nesvorny2023} and 0.0075~yr$^{-1}$ \citep{Zahnle2003}. On the other hand our impact probabilities with the satellites scaled to the giant planets are consistent with these earlier literature estimates, as is the leakage rate of objects from beyond Neptune into the Centaur region. However, our absolute impact probabilities with the giant planets are lower. We attribute this to our choice of initial conditions.}
{We present the current impact rates onto the regular satellites of Jupiter, Saturn and Uranus. We argue that our lower impact rate compared to earlier literature estimates is due to basing our results on the flux of objects coming in from beyond Neptune rather than relying on the current (observed) impact rate with Jupiter. We stress the importance of clearly stating all parameters and assumptions in future studies to enable meaningful comparisons.}


\maketitle
%
\section{Introduction}
The Jovian satellite Europa and the Saturnian satellite Enceladus are attractive astrobiological targets due to their activity, subsurface oceans and mostly young surface ages derived from a lack of impact craters. Yet any meaningful surface age for these bodies has to rely on an accurate derivation of the current impact rate on these bodies. \\

\citet{Zahnle1998} and \citet{Zahnle2003} used the numerical simulations of the outer solar system of \citet{LevisonDuncan1997} and a Monte Carlo method to compute the current cratering rates of heliocentric planetesimals on the regular satellites of the giant planets. \citet{Zahnle2003} presented two cases, one based on the abundance of small objects near Jupiter (Case A), and another from the impact crater size-frequency distribution on Triton (Case B). These works assumed that the total number of planetesimals crossing the orbits of the giant planets declined with time as $t^{-1}$ \citep{HolmanWisdom1993}. With this assumption the heavily cratered regions of most of the regular icy satellites were computed to be older than 4~Ga, and often as old as the Solar System itself when the Case A was considered, but much younger under the scenario Case B. However, the decline in impact flux does not evolve as $t^{-1}$ but as a stretched exponential \citet{Wong2020}, which can substantially modify the ages, particularly for old surfaces. As such, from these two studies alone there is a large variety in the impact flux onto these satellites and the resulting surface ages.\\

A more recent attempt to calculate the cratering rate on the regular satellites of the giant planets was undertaken by \citet{Nesvorny2023}. They rely on the extensive numerical simulations of the outer solar system published in \citet{Nesvorny2017b}, and they find similar impact probabilities and impact velocities of heliocentric bodies with the satellites compared with \citet{Zahnle2003}. Both \citet{Nesvorny2023} and \citet{Zahnle2003} scale the impact rates and impact probabilities of bodies with the satellites to that of Jupiter. \\

This study is our attempt to compute the current cratering rate based on our own simulations. It builds on the previous studies by \citet{Wong2019, Wong2020,Wong2023} and ongoing attempts to compute the surface ages of Enceladus. Unlike most previous studies we follow \citet{Levison2000a} and base our current impact rate onto the giant planets and their satellites from the leakage rate of scattered disc objects beyond Neptune into the Centaur region.

\section{Methodology and existing data}
There are several different methodologies in the established literature to compute the current cratering rate on the icy satellites. This work describes our approach to the problem. We shall also give a few comparisons to previous works. 

\subsection{Database of numerical N-body simulations of the outer solar system}
The cratering rate of the satellites of the giant planets relies on the dynamical evolution of the outer solar system. For this work we rely on a database of simulations that were reported on in \citet{Wong2019,Wong2020,Wong2023}.\\ 

We have run two sets of simulations of giant planet migration. The first set, reported in \citet{Wong2019}, employed the Graphical Processing Unit (GPU) integrator Gravitational ENcounters with GPU Acceleration (GENGA) \citep{GrimmStadel2014,Grimm2022} wherein the giant planets feel the gravitational force from the planetesimals, and the migration of the giant planets was the result of energy and angular momentum conservation as they scattered the planetesimals \citep{FernandezIp1984}. These simulations are denoted as {\it GPU M}. The second set with giant planet migration, reported in \citet{Wong2023}, used the Regularised Mixed Variable Symplectic (RMVS) integrator \citep{Levison1994}, version 4, wherein the migration of the planets was orderly due to the application of fictitious forces \citep{Levison2008}, and the planetesimals were modelled as massless test particles; this set is hereafter denoted as {\it RMVS M}. Both sets of migration simulations began with the giant planets on a more compact configuration surrounded by a disc of planetesimals. From the GENGA simulations we obtained the impact probability of a planetesimal with a giant planet, and we recorded the vectors of planetesimals one time step after their closest approach with a giant planet within 40 planetary radii ($R_{\rm P}$). From these vectors we enacted a high number of simulations wherein we subjected the regular satellites to the planetesimals that flew past them. We computed the impact probability of a planetesimal with a satellite from these flyby simulations and the encounter probability with the planets \citet{Wong2019}.\\

In \citet{Wong2023} we discovered that the RMVS migration simulations did not have enough resolution, i.e. remaining particles, to accurately compute the impact chronology onto the giant planets over the last 3.5~Gyr. We therefore ran additional simulations with RMVS and GENGA: we cloned all planetesimals after 1 Gyr of dynamical evolution after the controlled giant planet migration with RMVS. We ran these cloned planetesimals for another 3.5 Gyr to obtain a detailed impact chronology with high resolution data and robust impact statistics; these simulations are denoted as {\it RMVS C} and {\it GPU C} respectively. One additional GENGA simulation integrated all test particles that remained after 1 Gyr for another 3.5 Gyr without cloning (GPU Mixed). A summary of the different simulations are presented in Table~\ref{tab:sims}. From them we computed the impact probabilities ($P_{\rm P}$) with the giant planets, and from the GENGA simulations we obtained the encounter probabilities within 40 planetary radii ($P_{\rm enc}$) as well.\\

\begin{figure}
\centering
 \resizebox{\hsize}{!}{\includegraphics{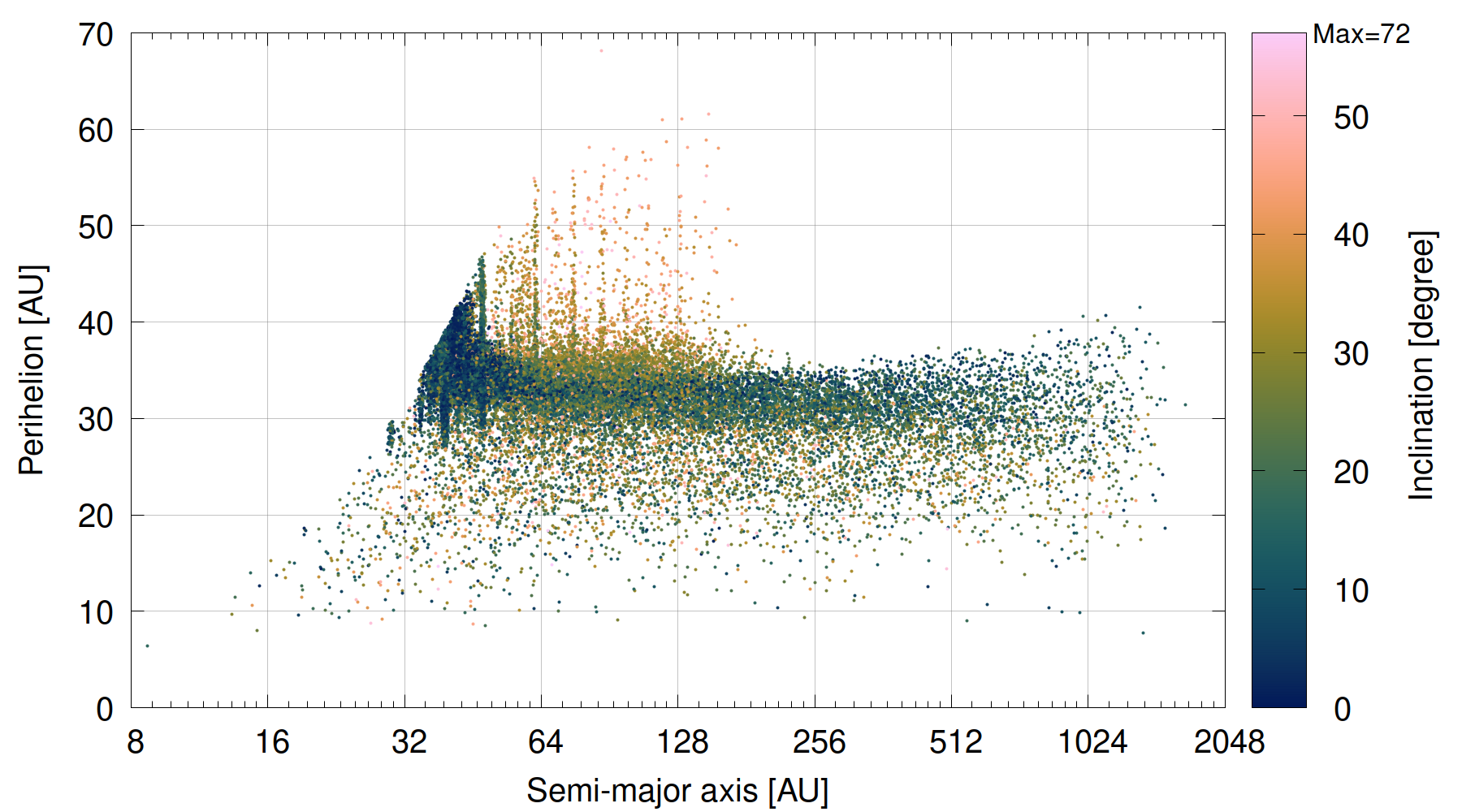}}
\caption{\label{fig:tp1g}Distribution of planetesimals after 1 Gyr of evolution of giant planet migration. Data from \citet{Wong2023}.}
\end{figure}

In what follows we shall consider the migration phase, i.e. the first 1 Gyr, to be separate from the last 3.5~Gyr, and we shall not discuss the migration phase because it is unimportant for our purpose: \citet{Nesvorny2023} state that 'the present properties of the ecliptic comet population are not sensitive to the details of Neptune’s early migration'. The reason for treating the last 3.5~Gyr differently is that in order to compute the current cratering rate onto the satellites we need to calculate the current injection rate of comets from beyond Neptune towards the giant planets. For this we need to simulate a sample of particles whose orbital distribution is similar to the unbiased observed population, or, in our case, the population of particles remaining after 4.5~Gyr of dynamical evolution. This sample will have a very different orbital distribution than the initial conditions of the migration simulations, and as such the impact and encounter probabilities with the giant planets of this population will be different than during the migration phase. In \citet{Wong2023} we used the state vectors of the RMVS M migration simulations after 1 Gyr as a proxy for the initial conditions of the 'current' population, and we build on those results here. The semi-major axis, perihelion distance and inclination of this population is shown in Figure~\ref{fig:tp1g}.

\begin{table}
    \centering
    \begin{tabular}{l|cccc}
    Model & GPU M & GPU C & RMVS M & RMVS C \\
    \cmidrule(lr){1-1} \cmidrule(lr){2-3} \cmidrule(lr){4-5} 
    Used & No & Yes & Yes & Yes \\ 
    $N_{\rm init}$ [$10^6$] & 1.0 & 0.037-0.08 & 2.05 & 1.31 \\
    Versions & - & c, m, e & c, e & c, e \\
    Step size [d] & 121.75 & 121.75 & 146.1 & 146. 1\\
    Output [Myr] & 0.01 & 0.3 & 0.1 & 0.1 \\
    Length [Gyr] & 0.2 & 3.5 & 4.5 & 3.5 \\ \\
    \end{tabular}
    \caption{In the models M stands for simulations including giant planet migration, C stands for clone i.e. post-migration. In the versions c = compact disc (outer edge at 31 au or 33 au), e = Extended disc (outer edge at 35, 37 or 39 au) and m = mixed disc (data from discs with all outer edges are combined). The m version of the GPU C simulation did not use particle cloning.}
    \label{tab:sims}
\end{table}
\subsection{Computing the satellite impact probabilities}
There are two ways to calculate the impact probabilities of the planetesimals with the satellites ($P_{\rm s}$). The first method makes use of the relative impact probabilities of the planetesimals on the satellites versus on the host planet. \citet{Zahnle2003} computes that
\begin{equation}
    \frac{P_{\rm s}}{P_{\rm P}} \approx \Bigl(\frac{R_{\rm s}}{a_{\rm s}}\Bigr)^2\frac{N(<a_{\rm s})}{N(<R_{\rm P})},
    \label{eq:pspp}
\end{equation}
where $R_{\rm s}$ is the satellite's radius, $a_{\rm s}$ is the  the semi-major axis of the satellites' orbits around the planet, and the remaining term is the ratio between the number of planetesimals that crossed the orbit of the satellites ($N(<a_{\rm s})$) and those that impact on the planet ($N(<R_{\rm P})$). Furthermore, the latter fraction in Eq.~\ref{eq:pspp} can be expressed as a function of the planetesimal's hyperbolic velocity 'at infinity' ($v_{\infty}$) and the satellites' orbit velocities ($v_{K}$) as \citep{Zahnle2003}
\begin{equation}
 \frac{N(<a_{\rm s})}{N(<R_{\rm P})} = \frac{v_{\infty}^2+2v_K^2}{v_{\infty}^2(R_{\rm P}/a_{\rm s})^2+2v_K^2(R_{\rm P}/a_{\rm s})}.
 \label{eq:nanr}
\end{equation}
It should be noted that $2v_K^2 = v_{\rm esc}^2$, where the latter is the escape speed at the distance of the satellite from the planet. Typically, from our GPU M simulations we find that $\langle v_{\infty} \rangle \sim (0.27-0.38) v_{\rm P}$, where $v_{\rm P}$ is the orbital velocity of the planet around the Sun; this agrees with earlier estimates by  \citet{Vokrouhlicky2008,Nesvorny2017b}. Therefore $v_{\infty} < v_K$ for the regular satellites except for Iapetus. The former expression can then be simplified to a linear relationship given by
\begin{equation}
\frac{N(<a_{\rm s})}{N(<R_{\rm P})} \sim \frac{a_{\rm s}}{R_{\rm P}}.
\label{eq:nanr2}
\end{equation} 
This approximation breaks down when $v_K \sim v_{\rm enc}$, which can be solved for the distance to the planet. With the typical encounter velocities that we obtain from the GPU M simulations for Jupiter this distance is approximately 110~$R_{\rm J}$, for Saturn it is 75~$R_{\rm S}$ and for Uranus this happens around 52~$R_{\rm U}$. At a greater distance $\frac{N(<a_{\rm s})}{N(<R_{\rm P})} \sim (\frac{a_{\rm s}}{R_{\rm P}})^2$. By substituting the linear relationship into Eq.~\ref{eq:pspp}, we have
\begin{equation}
\frac{P_{\rm s}}{P_{\rm P}} \sim \frac{R_{\rm s}}{a_{\rm s}}\frac{R_{\rm s}}{R_{\rm P}}.
  \label{eq:zahnle03b}
\end{equation}
Eq.~\ref{eq:nanr2} and \ref{eq:zahnle03b} reflect the intense gravitational focusing by the giant planets for close encounters.\\
 
For the second method we employ numerical flyby simulations as detailed in \citet{Wong2019}. We can express $P_{\rm s}$ as the fraction of planetesimals that collided on the satellites and multiply it by the probability of the planetesimals coming within 40~$R_{\rm P}$ of a giant planet; we call the latter the encounter probability ($P_{\rm enc}$). We have
\begin{equation}
    P_{\rm s} = \frac{N_{\rm i}}{N_{\rm tot}}\ P_{\rm enc}, 
\label{equ:psat}
\end{equation}
where $N_{\rm i}$ is the recorded number of impacts on the satellites and $N_{\rm tot}$ is the total number of planetesimals that come with a distance of 40~$R_{\rm P}$. From the flyby simulations of the three planet-satellites systems, we could count the number of impacts on each regular satellite and calculate the absolute impact probabilities.

\subsection{Re-enacting the close encounters}
For this work we have decided to redo the flyby simulations using the CPU version of GENGA. The reason is the following. Following \citet{Chambers1999}, the GENGA integrator computes a critical radius around each massive body in the simulation, whose magnitude is given by $R_{\rm c}=$max$(n_1R_H,n_2v_Kh$). Here $R_H$ is the Hill radius of the body, $h$ is the time step, and $n_1$ and $n_2$ are constants set by the user; their default values are 3.0 and 0.4. For example, for Mercury using a time step of $h=0.01$ yr we have $R_{\rm c}=0.04$~au which is dominated by the second term, but for Jupiter with $h=0.4$~yr we have $R_{\rm c}=1.07$~au which is dominated by the first term. Within this critical distance two bodies are considered to be within a close encounter, and the integration routine smoothly changes from the Mixed Variable Symplectic (MVS) map \citep{WH1991} beyond $R_{\rm c}$ to the Bulirsch-Stoer routine \citep{GrimmStadel2014,Grimm2022} when the two bodies are closer than 0.2$R_{\rm c}$. The changeover between the MVS and Bulirsch-Stoer regimes needs to be done smoothly so that during a close encounter the integrator needs to perform at least a few steps inside the changeover regime between 0.2$R_{\rm c}$ and 1$R_{\rm c}$.  \\

The default values of $n_1$ and $n_2$ of 3.0 and 0.4 suffice for many problems: the value of $n_1$ is identical to the default value in RMVS \citep{Levison1994} and is typically sufficient for low-velocity encounters, or when the mass ratio between the central body and the planets or satellites is greater than of the order of $10^{-5}$. However, when the masses are lower, the key distance factor is not the number of Hill radii, but rather the number of time steps used to sample the changeover region, i.e. the second term. A pair of objects with lunar mass can easily travel 3 mutual Hill radii in a single time-step. In this case the switch from the symplectic regime to the Bulirsch-Stoer regime would be instantaneous, leading to large errors during the close encounter. As such, for high-velocity encounters, the $n_2$ parameter becomes dominant \citep{Chambers1999}. \\

The average encounter speed between the regular satellites of the giant planets and the hyperbolic heliocentric planetesimals can be computed with Pythagoras' theorem and is approximately $\sqrt{3}v_K$ \citep{Zahnle2003}, so that within a time step the relative distance the satellite and a planetesimal travel is about $\sqrt{3}v_K h$. This distance is typically much greater than a few Hill radii of the satellites. As such, the $n_2$ term dominates, and it is likely that $n_2=0.4$ is too low and the simulations may 'step over' close encounters and collisions between the satellites and the planetesimals. As such, we have redone the simulations by setting $n_2=1$. We reckon that this value is high enough because GENGA begins to check for potential close encounters at a distance of $\sqrt{3}R_{\rm c}$.\\

Due to the low impact probabilities with the satellites it is necessary to re-enact many more encounters than are obtained during the giant planet migration simulations. We sampled the velocity distribution of the planetesimals as they penetrated the 40 planetary radii sphere, and recorded their values once they were between 38 and 42 planetary radii. We created 25\,000 massless planetesimals in this manner per simulation, and ran 20\,000 simulations per planet for 500 million planetesimals per planet; the maximum number of planetesimals per simulation was limited by computing time: higher numbers of planetesimals would increase the computing time more sharply than a linear increase due to greater RAM and CPU cache occupation. From testing we found that using 25\,000 planetesimals using one CPU core gave excellent results, with each simulation lasting 12-17 seconds. In these flyby simulations all planetesimals passed by the satellites at once.\\

The isotropic initial position vectors of the planetesimals are computed according to
\begin{eqnarray}
\cos \theta &=& 2\xi_1 -1, \nonumber \\
\sin \theta &=& \sqrt{1-\cos^2 \theta} \nonumber \\
\phi &=& 2\pi\xi_2, \nonumber \\
x &=& S \sin \theta \cos \phi, \nonumber \\ 
y &=& S \sin \theta \sin \phi, \nonumber \\
z &=& S \cos \theta,
\end{eqnarray}
where $\xi_1$ and $\xi_2$ are uniform random numbers in the interval $[0,1]$, and $S$ is the distance to the planet corresponding to 40~$R_{\rm P}$. The velocities of the planetesimals are computed from angular momentum conservation and the requirement that they are initially all moving towards the planet. Their components are computed as

\begin{eqnarray}
\psi &=& 2\pi\xi_3, \nonumber \\
v_3&=&-u\sqrt{\xi_4}, \nonumber \\
v_1 &=& \sqrt{u^2-v_3^2}\cos \psi, \nonumber \\
v_2 &=& \sqrt{u^2-v_3^2}\sin \psi, \nonumber \\
v_x&=& -v_1\sin \phi +v_2\cos \theta \cos \phi + v_3\sin \theta \cos \phi, \nonumber \\
v_y&=& v_1 \cos \phi + v_2 \cos \theta \sin \phi + v_3 \sin \theta \sin \phi, \nonumber \\
v_z&=& -v_2 \sin \theta + v_3 \cos \theta,
\end{eqnarray}
where $\xi_3$ and $\xi_4$ are two additional uniform random numbers, and $u$ is the speed of the particle at 40~$R_{\rm P}$, whose magnitude and distribution was computed from the GENGA simulations. With this prescription, the impact parameter of the planetesimals, $b=L/u$, where $L$ is the specific angular momentum, is uniformly distributed in $b^2$. Even though this impact parameter distribution is only strictly valid beyond about 52~$R_{\rm U}$ for Uranus, 75~$R_{\rm S}$ for Saturn and beyond about 110$~R_{\rm J}$ for Jupiter, this prescription does reproduce the trend from the GENGA planet migration simulations that $P_{\rm enc}/P_{\rm P} \sim 50$ found by \citet{Wong2019}. We computed the hyperbolic velocities at 40~$R_{\rm P}$ from the migration and post-migration GENGA simulations discussed in \citet{Wong2023}, and from there calculated the encounter velocities 'at infinity' as $v_\infty = \sqrt{u^2-v_{\rm esc}(r=40\,R_{\rm P})^2}$. For Jupiter $\langle u \rangle = 10.1$~km~s$^{-1}$ leading to $\langle v_\infty \rangle = 3.17$~km~s$^{-1}$, for Saturn $\langle u \rangle = 6.85$~km~s$^{-1}$ leading to $\langle v_\infty \rangle=3.65$~km~s$^{-1}$ and for Uranus $\langle u \rangle = 4.25$~km~s$^{-1}$ leading to $\langle v_\infty \rangle=2.39$~km~s$^{-1}$. \\

For the purpose of our calculations, we assumed that the satellites reside on their present-day orbits. The initial positions and velocities of the Jovian, Saturnian and Uranian satellites were obtained from the JPL Horizons website. The time step was 0.01 days and the maximum simulation time was just 20 days, by which time most of the planetesimals had left. For those few planetesimals that impacted a satellite their impact velocities were recorded. 

\section{Results and outcomes}
\subsection{Fundamental simulation parameters}
\label{sub:nec}
Many small bodies in the trans-Neptunian region of the solar system are on meta-stable orbits. These orbits could appear to be stable for billions of years because they are temporarily trapped in mean-motion resonances with Neptune \citep{Duncan1995,DuncanLevison1997}. Non-resonant objects with long semi-major axes whose perihelion lies within $\sim$37~au evolve on billion-year timescales, undergoing a random walk in semi-major axis \citep{Fernandez2004}, while resonant objects, once dislodged, can become Neptune crossing \citep{Duncan1995,DuncanLevison1997}. At low semi-major axis (often $\lesssim 100$~au) secular effects are important, which alter the eccentricity and thus the perihelion distance. These secular effects can flip a body from non-crossing to a crossing orbit, changing them from a scattered disc object to a Centaur or ecliptic comet. These combined dynamical effects cause a slow leakage of scattered disc and resonant objects into the Centaur region.\\

In order to compute the current cratering rate onto the regular satellites we need to rely on the distribution of ecliptic comets, and more specifically on the injection rate of scattered disc objects into the realm of the giant planets. Ecliptic comets are scattered disc objects that have left their source region and become Neptune crossers; their motion is thus dominated by encounters with the planets and secular oscillations in inclination and eccentricity. A subset of these comets are the short-period Jupiter-family comets, which are concentrated along the ecliptic plane, and whose Tisserand's parameter with respect to Jupiter $T>2$ \citep{LevisonDuncan1997}. The Tisserand parameter with respect to a planet is akin to the Jacobi integral in the circular restricted three-body problem, and is roughly conserved during encounters with said planet. We qualified the particles in our simulations that met all the following criteria as an ecliptic comet (EC). Following \citet{Wong2023} our definition focuses on the EC's injection rate, and not so much their actual fate. Our criteria are:
\begin{enumerate}
	\item the planetesimal survived the first 1 Gyr during the migration episode;
	\item its perihelion reached $q<30$ au in the last 3.5 Gyr of the simulation, and  
	\item it is lost eventually due to collision with the planets, ejection from the Solar System, or survived till the end of simulation at 4.5 Gyr.
\end{enumerate}
As such, only those scattered disc objects that began to cross Neptune are counted as ecliptic comets. \\
	
Following \citet{Duncan1995} we compute the injection rate of ecliptic comets from the scattered disc as
\begin{equation}
F_{\rm EC} = N_{\rm SD} \vert r_{\rm SD}\vert f_{\rm EC},
\label{eq:nec}
\end{equation}
where $N_{\rm SD}$ is the current number of objects in the scattered disc, $r_{\rm SD}$ is the rate at which the scattered disc population declines, and $f_{\rm EC}$ is the fraction of scattered disc objects that become ecliptic comets. The rate of decline of scattered disc objects is computed as the number of planetesimals that left the system ($\Delta N$) divided by the final number of the particles at the end ($N_{\rm fin}$) and the time interval ($\Delta t$)  \citep{Duncan1995}. In other words
\begin{equation}
	r_{\rm SD} = \frac{1}{N_{\rm fin}}\frac{\Delta N} {\Delta t}. 
\end{equation}
It is worth mentioning that $r_{\rm SD}$ depends on the time range over which it is calculated because the decline is not constant with time. Here, we calculated $\Delta N$ and $N_{\rm fin}$ over the last 0.5 Gyr in the simulations to approximate the current injection rate. This approach is not without problems, however. The quantity $\Delta N$ changes discretely, and if $\Delta N$ is a small quantity but $N_{\rm fin}$ is large, then the uncertainty in the ratio between different simulations will likely be low. But if $N_{\rm fin}$ is also a small quantity, then the ratio $\Delta N/N_{\rm fin}$ can take on an (arbitrary) large range in values due to stochastic variations in $\Delta N$ between different simulations. As such, the expectation is that the uncertainties in $r_{\rm SD}$ will become large when the number of remaining particles in the simulations is low, and these uncertainties will not be reduced by running more simulations with the same initial number of particles. \\

To estimate the total number of ecliptic comets we need their mean dynamical lifetime, $\tau_{\rm EC}$. \cite{Levison2000a} argue that 
\begin{equation}
	\tau_{\rm EC} \approx \int_{0}^t \frac{N_{(s)}}{N_{\rm init}}ds \approx \frac{1}{N_{\rm init}} \sum_{i=1}{N_{\rm init}} t_i,
\end{equation}
where $N_{\rm init}$ is the initial number of test particles and $N_{(s)}$ is the number of test particles remaining at the time $s<t$. The last step is approximately valid because of the slow decline of planetesimals. The orbital elements or vectors of each body during the simulation are written to disk at regular time intervals: every 0.1~Myr for RMVS and every 0.3~Myr for GENGA. In this manner each simulation produces thousands of data data frames. To compute $\tau_{\rm EC}$ we, (a) searched each data frame for a particular particle, (b) determined whether its perihelion $q<30$~AU, and (c) counted the number of data frames the particle was in with $q<30$~au. If its perihelion increased again beyond 30~au it was not counted. We computed $\tau_{\rm EC}$ for each particle by multiplying the total number of data frames by the data output interval (either 0.1~Myr or 0.3~Myr). We then averaged these individual times to compute $\tau_{\rm EC}$. \\

The nominal average values of all of these quantities are listed in Table~\ref{tab:disc} for the different sets of numerical simulations that we performed. All the quantities derived from the simulations, as well as their uncertainties, are in the Supplementary Materials. We can then crudely average the results over all the simulations. In what follows we list all the relevant quantities from all simulation sets, but for the analysis of the impact rate we shall only use the clone (C) simulations. \\

We compute $r_{\rm SD}=-0.108^{+0.225}_{-0.108}$ Gyr$^{-1}$ (2$\sigma$) from the RMVS M simulations, $r_{\rm SD}=-0.179^{+0.030}_{-0.040}$ Gyr$^{-1}$ (2$\sigma$) from the RMVS C simulations, and $r_{\rm SD}=-0.149 \pm 0.004$ Gyr$^{-1}$ (1$\sigma$) from the the combined GPU simulations. 
Most of our values are in between those of previous literature estimates and their uncertainties from \citet{Brasser2013} ($-0.163\pm 0.66$), \citet{Fernandez2004} (-0.15), \citet{Levison2006} (-0.27), \citet{VolkMalhotra2008} ($-0.15\pm 0.05$), but are much lower than that of \citet{DiSisto2007} (-0.52). 
Some of these studies \citep{Fernandez2004,DiSisto2007,VolkMalhotra2008} computed the injection rate from integrating the observed sample of then-known trans-Neptunian objects, while others \citep{Levison2006,Brasser2013} relied on a distribution of planetesimals that resulted from long-term integrations of particles subjected to giant planet migration. The agreement between all these results is encouraging and suggested that the leakage rate may only be weakly dependent on the underlying orbital distribution of scattered disc objects.\\ 

From the RMVS M simulations we calculated $f_{\rm EC} = 66.6^{+22.8}_{-21.2}\%$ (2$\sigma$), $f_{\rm EC}=68.1^{+20.0}_{-5.2}\%$ (2$\sigma$) from the RMVS C simulations, and $f_{\rm EC} = 78.7\% \pm 1.2\%$ (1$\sigma$) from the GPU runs. The mean lifetime of the ecliptic comets in the RMVS migration simulations is $\tau_{\rm EC} = 256^{+998}_{-253}$ Myr (2$\sigma$). The RMVS clone simulations yield $\tau_{\rm EC} = 167^{+423}_{-164}$ Myr (2$\sigma$) and the GENGA simulations result in $\tau_{\rm EC} = 154^{+379}_{-151}$ Myr (2$\sigma$). All of these timescales are comparable with that found by \citet{Levison2000a} of $\tau_{\rm EC}=190$~Myr, but longer than found by \citet{DiSisto2007} (72~Myr) and \citet{Tiscareno2003} (9~Myr), but the last study only simulated ecliptic comets whose perihelia were close to Jupiter and Saturn, drastically reducing the dynamical lifetime. \\

For completion we report that the total depletion for the RMVS M simulations is $99.63^{+0.48}_{-0.28}\%$ (2$\sigma$), for the RMVS C simulations it is $99.81^{+0.17}_{-0.11}\%$ (2$\sigma$) and $99.85\% \pm 0.05\%$ (1$\sigma$) for the GPU C simulations. For the RMVS M simulations the large uncertainties are caused by the low number of surviving planetesimals (of the order of 20), so that $\Delta N/N_{\rm fin}$ varies greatly between different simulations due to stochastic leakage and ejection. Last, we report the fraction of the total population in ecliptic comets with $q<30$~au, which is computed as $N_{\rm EC}=N_{\rm SD} \vert r_{\rm SD}\vert f_{\rm EC}\tau_{\rm EC}$. The mean and uncertainties were computed using a Monte Carlo sampling procedure from the underlying distributions in $f_{\rm EC}$, $r_{\rm SD}$ and $\tau_{\rm EC}$ across all simulations in each set. For the RMVS M simulations we obtain $N_{\rm EC}=0.054^{+0.218}_{-0.054}N_{\rm SD}$ (2$\sigma$), $N_{\rm EC}=0.024^{+0.061}_{-0.023}N_{\rm SD}$ (2$\sigma$) from the RMVS C simulations, and $N_{\rm EC}=0.018^{+0.027}_{-0.017}N_{\rm SD}$ (1$\sigma$) from the GPU simulations. The large uncertainties are almost entirely due to the distribution in $\tau_{\rm EC}$.\\

Before we can calculate the current impact rate onto the giant planets and their satellites we need to know the number of scattered disc objects greater than some threshold diameter, $N_{\rm SD}$.

\begin{table}
\centering
\begin{tabular}{lcccc} 
\hline \\
Model & $\vert r_{\rm SD}\vert$  & $f_{\rm EC}$  & $\tau_{\rm EC}$ & Depletion \\ 
\footnotesize{Unit} &\footnotesize{Gyr$^{-1}$} & \footnotesize{$\%$} &	\footnotesize{Myr} & \footnotesize{$\%$} \\ 
\cmidrule(lr){1-1}  \cmidrule(lr){2-5} 
Compact RMVS M  & 0.164 & 82.9 & 273 & 99.89 \\ 
Extended RMVS M& 0.071 & 55.8 & 250 & 99.47 \\ 
Average RMVS M& 0.108 & 66.6 & 256 & 99.63 \\ \\
Compact RMVS C  & 0.185 & 72.0 & 205 & 99.91 \\
Extended RMVS C& 0.173 & 64.2 & 124 & 99.75 \\ 
Average RMVS C& 0.179 & 68.1 & 167 & 99.81 \\ \\
Compact Clone GPU & 0.144 & 77.4 & 187 & 99.91 \\
Extended Clone GPU & 0.151 & 78.6 & 125 & 99.78 \\
Mixed disc no Clone GPU & 0.153 & 80.2 & 141 & 99.85 \\ 
Average Clone GPU & 0.149 & 78.7 & 154 & 99.85 \\
\end{tabular}
\caption{\label{tab:disc}Average quantities derived from the results of the numerical simulations with compact and extended discs. Columns 2 to 5 show the rate of decline of scattered disc objects ($r_{\rm SD}$), the fraction of scattered disc objects that become ecliptic comets ($f_{\rm EC}$), the dynamical lifetime of the ecliptic comets ($\tau_{\rm EC}$), and the percentage of total depletion. Uncertainties are in the main text and the Supplementary Materials.}  
\end{table}

\subsection{The current number of scattered disc objects and their size-frequency distribution}
\label{sub:nsd}
Based on Outer Solar System Origins Survey (OSSOS) observations the current estimated number of scattered disc objects with diameter $D_{\rm i}>10$~km is $(2.0 \pm 0.8) \times 10^7$ \citep{Nesvorny2019}, assuming their albedo is 6\%. We scaled the number of impactors from the observed number of scattered disc objects with $D_{\rm i}$. We have, 
\begin{equation}
\label{eq:n_sd_d}
N_{\rm SD,~>D_{\rm i}} =  N_{\rm scale} \left(\frac{D_{\rm i}}{10~\rm km}\right)^{\alpha}.
\end{equation}
Observations of distant scattered disc objects in the hot/excited population, as well as Jupiter's Trojan asteroids, show that the slope $\alpha$ seems to have two values, one for objects with diameter $D_{\rm i} \gtrsim 100$~km, and another one for smaller objects \citep{Shankman2013,Fraser2014,Lawler2018,WongBrown2015,YoshidaTerai2017}. These studies suggest that for $D_{\rm i}\gtrsim 100$~km $|\alpha_1| = 4.5 \pm 0.5$ and for $1\lesssim D_{\rm i} \lesssim 100$~km objects $|\alpha_2| = 2.0 \pm 0.2$. \citet{Wong2023} adopted the nominal values $\alpha_2=-2$ and $\alpha_1=-4.5$. For impactors with diameter $D_{\rm i}<1$~km, we shall adopt the slope of \citet{Singer2019}, i.e. $\alpha_3=-0.7$.\\

With the nominal adopted value of $\alpha_2$ by \citet{Wong2023} the expected current number of scattered disc objects with diameter $D_{\rm i}>1$~km is approximately $2.0\times10^9$, which is consistent within uncertainties with the values obtained by previous numerical and observational works \citep[e.g.][]{Brasser2013,VolkMalhotra2008,Shankman2013,Lawler2018}. This estimate varies by {\it at least} a factor of three: \citet{Shankman2013} computed $2\times 10^9$ for $D_{\rm i}>1.5$~km (assuming an albedo of 4\%), followed by \citet{Lawler2018} which increased it to $3\times 10^9$ for the same diameter. \citet{VolkMalhotra2008} report a number of $10^9$, probably for objects with $D_{\rm i}>2$~km (the exact diameter is not specified), and \citet{Brasser2013} estimate $2\times 10^9$ for $D_{\rm i}>2.3$~km. It is possible that these studies might have adopted different size-frequency distributions at that size range, making scaling all of these values to $D_{\rm i}>1$~km or $D_{\rm i}>10$~km challenging. Can we verify whether any of the above estimates for $N_{\rm SD}$ make sense? \\


\begin{table*}
    \centering
    \begin{tabular}{l|ccccccc}
         Parameter & |$r_{\rm SD}$| & $f_{\rm EC}$ & Depl. & $(|\alpha_2|,|\alpha_3|)$ & $N_{\rm SD}$ & $F_{\rm EC}$ & $(\rho_i, \rho_s)$\\
         \footnotesize{Unit} & \footnotesize{Gyr$^{-1}$} & \footnotesize{\%} & \footnotesize{\%} & \footnotesize{-} & \footnotesize{$10^9$} &
         \footnotesize{yr$^{-1}$} & \footnotesize{kg m$^{-3}$} \\ \hline
         Value & 0.15 & 79 & 99.8 & (2.1,0.7) & 3.0 & 0.36 & (400,1000)
    \end{tabular}
    \caption{The nominal adopted values of several relevant parameters.}
     \label{tab:nominal}
\end{table*}

One source of bodies that we can use are the Jupiter-family comets (JFCs). Following \citet{LevisonDuncan1997} can define the production rate of JFCs as $r_{\rm SD}f_{\rm JFC}$. Literature estimates show that $f_{\rm JFC} =16\%-30\%$ \citep{LevisonDuncan1997,DiSisto2009,Brasser2013}. The usually reported number of active JFCs with diameter $D_{\rm JFC}>2$~km and perihelion distance $q<2.5$~au is 117 \citep{DiSisto2009}. The dormant to active JFC ratio is about 5-6.5 \citep{LevisonDuncan1997,DiSisto2009,Brasser2013}, so to first order there are of the order of 700 dormant JFCs. These comets spend 7\%-10\% of their dynamical life with $q<2.5$ au \citep{Brasser2013}, so that there are about $10^4$ JFCs with $D_{\rm JFC}>2$~km in total. The median dynamical lifetime of these comets is 165~kyr \citep{LevisonDuncan1997,Brasser2013}, so the loss rate is then approximately 0.06~yr$^{-1}$. This must match the production rate $N_{\rm SD}f_{\rm JFC}|r_{\rm SD}|$ and results in a nominal value of $N_{\rm SD}=2\times 10^9$ if we take $f_{\rm JFC}=20\%$ and $|r_{\rm SD}|=0.15$~Gyr$^{-1}$, with a range of $(1.1-2.5)\times10^9$ accounting for the uncertainties of $f_{\rm JFC}$ and $|r_{\rm SD}|$. \\

From their own analysis of numerical simulations of JFC production, \citet{Nesvorny2017b} estimate that at present the number of active JFCs to scattered disc objects is $10^{-7}$ -- comparable to $6.8\times10^{-8}$ found by \citet{Brasser2013} -- and they compute that $N_{\rm SD}=(1.5-4)\times10^7$ with $D_{\rm i}>10$~km, implying $N_{\rm SD}=(1.5-4)\times 10^9$ with $D_{\rm i}>$1~km if we adopt $\alpha_2=-2$, consistent on the lower end with earlier estimates. In an independent study on near-earth objects including active JFCs \citet{Bottke2002} obtains $N_{\rm SD} = (2.8 \pm 2.0)\times10^9$ objects beyond Neptune with $D_{\rm i}>1$~km. However, most of the estimates for $N_{\rm SD}$ derived from the JFC population pertain to objects with diameter $D_{\rm i}>2$~km rather than 1~km, which agrees better with the estimates of \citet{Brasser2013,Shankman2013,Lawler2018} than with that calculated by us from the observational constraints by \citet{Nesvorny2019} and the numerical work by \citet{Nesvorny2017b} for objects with $D_{\rm i} >10$~km. \\

A potential problem presents itself when we extrapolate the estimate of the current population back in time before the onset of giant planet migration. The average depletion of the population in the clone simulations is 99.8\%, comparable to the 99.7\% reported in \citet{Nesvorny2017b}. If the current population with $D_{\rm i}>1$~km is 2 billion, the original number was about $10^{12}$, which is higher than that computed by \citet{Nesvorny2023} and \citet{Wong2023} by a factor of 1.5, which is acceptable. But if the current population is 3 or 4 billion then it is difficult to reconcile with the original population unless its size-frequency distribution slopes are on the very steep end of what observational constraints would allow. \\

One way out of this dilemma is to accept that the current population of scattered disc objects with diameter $D_{\rm i}>10$~km is $(2 \pm 0.8)\times 10^7$ \citep{Nesvorny2019} so that with our nominal depletion the nominal primordial population of such objects is $10^{10}$ -- \citet{Nesvorny2017b} reports $(8\pm 3)\times10^9$. With a primordial disc mass of 18~$M_\oplus$ adopted in \citet{Wong2019,Wong2023} and adopting the knee size frequency distribution of trans-Neptunian objects from \citet{Fraser2014,Lawler2018} a primordial number of objects of $10^{10}$ with $D_{\rm i}>10$~km requires that the size-frequency distribution slopes are either $(\alpha_1,\alpha_2)=(5.0,2.1)$ or $(\alpha_1,\alpha_2)=(4.5,2.2)$. In either case one of the slopes is at a maximum value consistent with observations. \\

With these faint-end slopes and our depletion amount the current number of scattered disc objects with $D_{\rm i}>1$~km becomes $N_{\rm SD}=(2.5-3.2)\times10^9$. In what follows we shall adopt a nominal value of $N_{\rm SD}=3\times10^9$ for $D_{\rm i}>1$~km. The resulting nominal number of ecliptic comets with $D_{\rm i}>1$~km is then $N_{\rm EC}=5.4\times10^7$ if we use the outcome of the GENGA simulations.\\

We can now compute the ecliptic comet injection rate. From the RMVS migration simulations we obtain $F_{\rm EC} = 0.216^{+0.499}_{-0.215}$~yr$^{-1}$ (2$\sigma$), for the RMVS clone simulations $F_{\rm EC} = 0.366^{+0.118}_{-0.082}$~yr$^{-1}$ (2$\sigma$), while the combined GENGA simulations yield $F_{\rm EC} = 0.353 \pm 0.002$~yr$^{-1}$ (1$\sigma$); these estimates do not include the uncertainty in $N_{\rm SD}$. We shall resort to using the outcome of the GPU clone simulations, and we adopt a nominal value of $F_{\rm EC} = 0.36$~yr$^{-1}$. With this approach the largest uncertainty in the cratering rate is that from the number of scattered disc objects.\\

In Table~\ref{tab:nominal} we have listed the nominal values of all the relevant parameters that we adopted.

\subsection{Impact probabilities with the giant planets and their satellites}

\begin{table*}[ht]
    \centering
    \begin{tabular}{l|ccc|ccc}
        Planet & $P_{\rm P}$ [GPU C] & $P_{\rm enc}$ [GPU C] & $P_{\rm P}$ [RMVS C] & $P_{\rm P}$ [GPU M] & $P_{\rm enc}$ [GPU M] & $P_{\rm P}$ [RMVS M] \\ 
        \footnotesize{Unit} & \footnotesize{\%} & \footnotesize{\%} & \footnotesize{\%} & \footnotesize{\%} & \footnotesize{\%} & \footnotesize{\%} \\ \hline
        Jupiter & 0.29 & 15.6 & 0.37 & 0.78 & 33 & 1.2 \\ 
        Saturn & 0.12 & 5.9 & 0.13 & 0.28 & 14 & 0.46 \\ 
        Uranus & 0.059 & 3.4 & 0.071 & 0.56 & 19 & 0.19
    \end{tabular}
     \caption{Here we list the impact and encounter probabilities, averaged for the various simulations. The C stands for the clone simulations, run for the last 3.5~Gyr, while M stands for the migration simulations, which were run either for 200~Myr (GPU) or 4.5~Gyr (RMVS).
     The impact probability of planetesimals with the planets during flyby is 1.6\% for Uranus, 1.9\% for Saturn and 2.1\% for Jupiter, close to the theoretical 2.5\%. Uncertainties are listed in the Supplementary Material.}
     \label{tab:pimpgp}
\end{table*}

\begin{table*}[ht]
    \centering
    \begin{tabular}{lc|ccccccccccccc}
        System & Satellite & $P_{\rm s}/P_{\rm P}$ & $N_i/N_{\rm tot}$ & $P_{\rm s}$ & $v_{\rm s}$ & $\dot{C}_{\rm s}$ & $D_{\rm cr}$ & $\dot{c}_{D>10}$\\ 
        & \footnotesize{Unit} & \footnotesize{$10^{-5}$} & \footnotesize{$10^{-6}$} & \footnotesize{$10^{-8}$} & \footnotesize{km s$^{-1}$} & \footnotesize{Gyr$^{-1}$} & \footnotesize{km} & \footnotesize{$10^{-6}$ km$^{-2}$ Gyr$^{-1}$}\\ \hline
        Jovian & Io & 13.5 & 2.85 & 44.4 & 32.0 & 160 & 8.3 & 2.5\\ 
        & Europa & 5.98 & 1.26 & 19.7 & 24.9 & 71.0 & 16.8 & 3.4\\ 
        & Ganymede & 11.6 & 2.45 & 38.2 & 20.4 & 138 & 14.9 & 2.1\\ 
        & Callisto & 5.90 & 1.24 & 19.4 & 15.4 & 70 & 13.0 & 1.2\\ \\
        Saturnian & Mimas & 0.58 & 0.11 & 0.63 & 23.2 & 2.26 & 26.4 & 10\\ 
        & Enceladus & 0.56 & 0.10 & 0.60 & 21.7 & 2.17 & 22.3 & 5.0\\ 
        & Tethys & 1.82 & 0.33 & 1.97 & 21.0 & 7.11 & 20.4 & 3.5\\ 
        & Dione & 2.72 & 0.50 & 2.96 & 18.8 & 10.6 & 17.1 & 4.1\\ 
        & Rhea & 2.45 & 0.45 & 2.66 & 15.1 & 9.58 & 14.7 & 1.8\\ 
        & Titan & 14.2 & 2.60 & 15.4 & 10.8 & 55.4 & 11.0 & 0.7\\ \\
        Uranian & Miranda & 2.77 & 0.44 & 1.46 & 12.4 & 5.25 & 18.2 & 12\\ 
        & Ariel & 10.5 & 1.63 & 5.47 & 10.2 & 19.7 & 11.9 & 5.4\\ 
        & Umbriel & 7.99 & 1.26 & 4.21 & 8.7 & 15.2 & 11.4 & 4.0 \\ 
        & Titania & 9.38 & 1.48 & 4.94 & 7.0 & 17.8 & 9.6 & 2.1\\ 
        & Oberon& 6.34 & 1.00 & 3.34 & 6.1 & 12.0 & 8.9 & 1.3
    \end{tabular}
    \caption{The impact probability of planetesimals with the satellites. The first two columns are the impact probability scaled to that of the planet as computed by \citet{Zahnle2003} and by us using the flyby simulations. The third column lists the impact probability of a planetesimal with a satellite during a flyby. The fourth is the total impact probability of a planetesimal with a satellite during the last 3.5~Gyr of the dynamical simulations. The penultimate column lists the average impact speed of a planetesimal with a satellite and the final column lists the impact rate of objects with diameter $D_{\rm i}>1$~km in units of Gyr$^{-1}$. For this last column the average of the two impact probabilities was used.
    }
    \label{tab:pimps}
\end{table*}
The various impact probabilities of planetesimals with the giant planets and the satellites, as well as encounter probabilities with the planets, and impact speeds with the satellites are listed in Tables~\ref{tab:pimpgp} and~\ref{tab:pimps}. The quantities derived from the numerical simulations, as well as their uncertainties, are given in the Supplementary Materials. \\

The first table lists the impact and encounter probabilities of planetesimals with the giant planets. The first two entries are obtained from the clone simulations run with GENGA for the last 3.5~Gyr \citep{Wong2023}. The third column is $P_{\rm P}$ for the RMVS clone simulations. The next two columns are the same values as the first two but are obtained from \citet{Wong2020} when the giant planets were migrating. The last column is the impact probability obtained from the RMVS simulations including migration. \\

Two things are of note. First, the impact probabilities with the planets after migration during the last 3.5~Gyr are lower than that during the migration phase, typically by a factor of 3--4. The reason is caused by the different initial conditions of the two sets of simulations, as explained earlier. During the migration phase the planetesimals are excited by the migrating planets and are quickly scattered from one planet to another. During the last 3.5 Gyr most planetesimals have been either ejected or have collided with one of the giant planets, and the remaining planetesimals reside on meta-stable orbits \citet{DuncanLevison1997} so that their time-averaged orbital distribution is very different from the first 1~Gyr. \\

Second, the impact probabilities computed with RMVS are systematically higher than with GENGA. The reason for this is likely to be the time step: its value was chosen as a compromise between speed and accuracy, with the understanding that most planetesimals are ejected due to the cumulative effect of mostly distant encounters with the giant planets. However, with a large enough time step RMVS does not adequately resolve very close repeated encounters with the giant planets \citep{Levison1994}, while GENGA, using a Bulirsch-Stoer integrator close to the planet, likely does a better job of resolving such close encounters. The impact probability difference between RMVS and GENGA is a factor of 1-1.5, which will be propagated into the cratering rates. In what follows we shall adopt the impact probabilities from the GENGA simulations.\\

The next table contains the impact probabilities of planetesimals with the satellites. The first entry is the impact probability scaled to that of the planet obtained from our flyby simulations, where we counted the number of impacts on a satellite and divided by the number that hits the planet. The next entry is the impact probability of a planetesimal that ventures within 40~$R_{\rm P}$ as computed from our flyby simulations. The next entry, $P_{\rm s}$ is the absolute impact probability, and is computed as $N_i/N_{\rm tot} \times P_{\rm enc}$, where we used the $P_{\rm enc}$ values on the planets from the GENGA simulations during the last 3.5~Gyr. As an example, for Enceladus $N_i/N_{\rm tot} = 0.10\times 10^{-6}$ and $P_{\rm enc}=5.9\%$ for Saturn so that $P_s = 0.1\times 10^{-6} \times 0.059 =0.59 \times 10^{-8}$, which is a little lower than using $P_{\rm s}/P_{\rm P}\times P_{\rm P}$. The discrepancy between these two values of the absolute impact probability is largest for Mimas and Enceladus, and is probably caused by the flyby simulations missing encounters and impacts due to the high encounter speeds with these tiny bodies. We subsequently list the average impact speed with the satellite obtained from the flyby simulations, followed by the impact rate of objects with $D_{\rm i}>1$ km, and the crater diameter that such an impact produces. The last entry is cratering rate for craters with diameter $D_{\rm cr}>10$~km.\\

We note that the differences between $P_{\rm s}/P_{\rm P}$ and with direct hits during the flyby simulations are typically 30\% and are smallest for the Jovian satellites Io, Europa and Ganymede. We do not know what accounts for this large difference. Both the analytical estimate of \citet{Zahnle2003} and our numerical flyby simulations have their own assumptions and limitations, but \citet{Zahnle2003} did say that their Monte Carlo procedures gave results that were 30\% different from their analytical solutions. Additional differences could result from numerical factors in the flyby simulations and the fact that the gravitational focusing of the giant planets becomes important for Saturn and Uranus at closer distances than for Jupiter, resulting in a different initial distribution of the impact parameter. The uncertainty in the value of $P_{\rm P}$ is at least a factor of 1.3. 

\subsection{Current cratering rate on the satellites}
\label{sub:enc}
The current impact rate on a satellite (or a planet) is computed following the procedure from \citet{Duncan1995} and \citet{Levison2000a}, and it’s based on the injection rate ($F_{\rm EC}$) of ecliptic comets with $D_{\rm i} \geq 1$ km originating from beyond Neptune and the impact probability with the satellites and giant planets. The formula for calculating the impact rate is:
 \begin{equation}
     \dot{C}_{\rm s} = F_{\rm EC}~P_{\rm s}
 \end{equation}
where $P_{\rm s}$ is the post-migration collision probability onto a satellite as described in equation \ref{equ:psat}. \\

The impact rates listed in the fifth column of Table~\ref{tab:pimps} need to be converted into a cratering rate using crater scaling laws to convert impactor diameter to craters, and vice versa. \citet{Wong2023} employed a crater scaling law based on the $\Pi$-scaling law, using the fact that the simple-to-complex crater diameter $D_{\rm SC}=4~(1\,{\rm m\,s}^{-2}/g)$~km \citep{Schenk1991}, which applies to all icy satellites apart from Mimas, Enceladus, Tethys, Dione and Miranda, for which $D_{\rm SC} \sim 15$~km \citep{Schenk1991}. For Io, $D_{\rm SC}=24~(1\,{\rm m\,s}^{-2}/g)$~km \citep{Pike1980}. The crater scaling adopted in \citet{Wong2023} for icy bodies becomes

\begin{eqnarray}
D_{\rm cr}&=&4.66\left(\frac{D_{\rm i}}{1\,{\rm km}}\right)^{0.917} \left(\frac{\rho_i}{\rho_s}\right)^{0.392}\left(\frac{v_{\rm imp}}{1\,{\rm km\,s}^{-1}}\right)^{0.517}\left(\frac{g}{1\,{\rm m\,s}^{-2}}\right)^{-0.082} \nonumber \\
D_{\rm cr}&=&3.67\left(\frac{D_{\rm i}}{1\,{\rm km}}\right)^{0.917} \left(\frac{\rho_i}{\rho_s}\right)^{0.392}\left(\frac{v_{\rm imp}}{1\,{\rm km\,s}^{-1}}\right)^{0.517}\left(\frac{g}{1\,{\rm m\,s}^{-2}}\right)^{-0.259}
\end{eqnarray}
where the bottom equation should be applied to Mimas, Enceladus, Tethys, Dione and Miranda, and the top equation to the other satellites (for Io the coefficient in the top equation is 3.39). For both crater scaling laws we assumed that the impact angle is 45$^\circ$. \citet{Wong2023} assumed an impactor density of $\rho_{\rm i}= 400$~kg~m$^{-3}$, which is the mean density of comet 9P/Tempel 1 \citep{Richardson2007} and which is a little lower than that of 67P/C-G \citep{Jorda2016}, and a satellite density $\rho_{\rm s}= 1000$~kg~m$^{-3}$ apart from Io, for which $\rho_{\rm s}= 3520$~kg~m$^{-3}$ was assumed. The inverse relations for icy bodies are

\begin{eqnarray}
D_{\rm i}&=&0.187\left(\frac{D_{\rm cr}}{1\,{\rm km}}\right)^{1.09} \left(\frac{\rho_i}{\rho_s}\right)^{-0.427}\left(\frac{v_{\rm imp}}{1\,{\rm km\,s}^{-1}}\right)^{-0.564}\left(\frac{g}{1\,{\rm m\,s}^{-2}}\right)^{0.09} \nonumber \\
D_{\rm i}&=&0.241\left(\frac{D_{\rm cr}}{1\,{\rm km}}\right)^{1.09} \left(\frac{\rho_i}{\rho_s}\right)^{-0.427}\left(\frac{v_{\rm imp}}{1\,{\rm km\,s}^{-1}}\right)^{-0.564}\left(\frac{g}{1\,{\rm m\,s}^{-2}}\right)^{0.282},
\end{eqnarray}
from which we computed the cratering rate for $D_{\rm cr}>10$~km using the inferred impactor size-frequency distribution slope of $\alpha_3=-0.7$ for impactors with $D_{\rm i}<1$~km \citep{Singer2019}, and following \citet{Nesvorny2019} by adopting $\alpha_2=-2.1$ for $D_{\rm imp}>1$~km. The uncertainties in the cratering rate are a factors of a few, primarily due to uncertainties in $F_{\rm EC}$ (including in $N_{\rm SD}$). For Io the coefficient of the top equation becomes 0.264.\\

The nominal value of $F_{\rm EC}$ and the {\it current} impact probability of Jupiter gives a rough estimation of the impact rate on Jupiter, which we compute to be 1.2 $\times$ 10$^{-3}$ per year for objects with $D_{\rm i}>1$~km. Our estimated Jovian impact rate falls towards the low end, but remains consistent with the reported values from previous studies. The ranges and uncertainty of impact rates as reported by \citet{Zahnle2003} are discussed in section~\ref{sub:earlies_works} but suffice to say that the range of these impact rates for $D_{\rm i}>1$~km spans two orders of magnitude, and the uncertainties in these estimates also vary. We can then also calculate the impact and cratering rates on Enceladus (and the other satellites of all three planets). The impact rate of objects with $D_{\rm i}>1$~km with Enceladus is computed to  be 2.12~Gyr$^{-1}$. For the cratering rate we use the inverse crater scaling law. With our adopted values for the impactor density, satellite density and impact velocity from Table~\ref{tab:pimps}, a crater with diameter $D_{\rm cr}>10$~km on Enceladus requires a projectile diameter of $D_{\rm i}=0.41$~km and adopting the size-frequency distribution slope $\alpha_3=-0.7$ from \citet{Singer2019} the current cratering rate becomes $4.9\times10^{-6}$~km$^{-2}$~Gyr$^{-1}$.

\section{Comparison with earlier work}
\label{sub:earlies_works}
We present here the three sets of studies that estimate current cratering rates and surface age for the icy satellites. These sets include:
\begin{enumerate}
    \item Classic studies \citep{Zahnle1998,Zahnle2003,Kirchoff2009},
    \item Recent studies by others \citep{Nesvorny2019,Nesvorny2023,Bottke2023,Bottke2024}, and
    \item Recent studies by us \citep{Wong2019,Wong2020,Wong2023}, and this work. 
\end{enumerate}
Each group of studies represents a series of papers that begin with planetary dynamics and impact rate estimations, progressing towards determining crater retention ages for the icy satellites. These works employed various methodologies and sources, ranging from numerical simulations of planetary migration to observational crater counts. It is important to note that no single paper in any of these sets comprehensively addresses the entire spectrum from planetary dynamics to geological history.\\

We shall discuss the differences in results from the three studies by computing/listing the impact rates on Jupiter and Enceladus, and the cratering rate on the latter, in Table \ref{tab:rates}. In the following, we address the large uncertainties and differences between our cratering rate estimates and those from previous studies.

\begin{table*}[ht]
    \centering
    \begin{tabular}{ll|ccccccc} 
        \hline\\
        \multicolumn{2}{l|}{Model} & \multicolumn{2}{c}{Classic Studies} & \multicolumn{2}{c}{Recent Studies} & \multicolumn{3}{c}{Our Studies} \\
        \multicolumn{2}{l|}{} & \footnotesize{Case A} & \footnotesize{Case B} & \footnotesize{No Disruption} & \footnotesize{With Disruption} & \footnotesize{Averaged} & \footnotesize{Compact} & \footnotesize{Extended}\\
        \cmidrule(lr){1-2} \cmidrule(lr){3-4} \cmidrule(lr){5-6} \cmidrule(lr){7-9}\\
        $\dot{C}_{\rm Jup}$ & \multirow{3}{*}{$D_\text{i} \geq 1$ km} & 7.50 & 26.1 & 
            4.37 & 3.19 & 1.04 & 0.81 & 1.24\\ 
        $\dot{C}_{\rm Sat}$ & & 3.15 & 11.0 & 
            1.41 & 1.31 & 0.42 & 0.34 & 0.48\\
        $\dot{C}_{\rm Ura}$ & & 1.88 & 6.52 & 
            1.25 & 1.22 & 0.21 & 0.13 & 0.18\\ \\
        \multirow{2}{*}{$\dot{C}_{\rm s}$} 
        & $D_\text{i} \geq 1$ km & 16.5 & 57.4 & 
            6.56 & 6.12 & 2.13 & 1.56 & 2.52\\
        & $D_\text{cr} \geq 10$ km  & 60.8 & 450 & 
            24.1 & 22.5 & 4.90 & 3.59  & 5.82 \\ \\
    \end{tabular}
    \caption{The current impact rate on the three giant planets: Jupiter ($\dot{c}_{\rm Jup}$), Saturn ($\dot{c}_{\rm Sat}$), Uranus ($\dot{c}_{\rm Ura}$) per thousand years (kyr$^{-1}$) and the impact rate on Enceladus ($\dot{c}_{\rm s}$) per billion years (Gyr$^{-1}$) while the unit for the cratering rate is per billion years per million square kilometre ($10^{-6}$~km$^{-2}$~Gyr$^{-1}$), as derived from dynamical simulations and size-frequency distributions from three groups of studies. The cratering rates were derived from the output of the GENGA simulations.}
    \label{tab:rates}
\end{table*}

\subsection{Classic studies}
\label{sub:Classic}
\citet{Zahnle1998,Zahnle2003} constrained the current impact rate on Jupiter through multiple measurements. \citet{Zahnle2003} reported an impact rate of approximately 4 $\times$ 10$^{-3}$ yr$^{-1}$ based on six encounters (smallest $\sim$ 1 km in diameter) within four Jovian radii over $\sim$350 years. \cite{Lamy2004} observed nine comets crossing Callisto's orbit over approximately 50 years, suggesting an impact rate of about 1.75 $\times$ 10$^{-3}$ per year for $D_{\rm i}>1$~km. Other estimations, using recorded collisions and close encounters, measurements of excess carbon monoxide content, crater counting on Ganymede, and inference from Near Earth Objects, range from 0.4 to 44 $\times$ 10$^{-3}$ impacts per year. The ranges \citet{Zahnle2003} reported are: historical records ($4.3 \times 10^{-3}$ to $2.3 \times 10^{-2}$ impacts per year), observed impact ($1.4 \times 10^{-2}$ to $4.4\times 10^{-2}$ yr$^{-1}$), excess carbon monoxide content ($7.7 \times 10^{-3}$ yr$^{-1}$), crater counting on heavily cratered terrain on Ganymede ($2.0 \times 10^{-3}$ to $4.8 \times 10^{-2}$ yr$^{-1}$), and inferred from Near Earth Objects ($1.3 \times 10^{-3}$ yr$^{-1}$). All impact rates listed above are scaled to impactors with diameters $D_{\rm i}>1$~km, assuming a cumulative power-law slope of $\alpha_2 = -2$ \citep{Singer2019}. In any case, the impact rate on Jupiter estimated by \citet{Zahnle1998,Zahnle2003} have uncertainties that vary significantly, ranging from a factor of three to an order of magnitude. \citet{Zahnle2003} calibrates the impact rate on Jupiter at $5^{+6}_{-3}\times 10^{-3}$ yr$^{-1}$ for impactors with $D_{\rm i}>1.5$~km.\\

This commonly cited Jovian impact rate for $D_{\rm i} \geq  1.5$ km, when translated using their Case A size-frequency distribution, corresponds nominally to $10^{-4}$~yr$^{-1}$ for impactors with diameter $D_{\rm i} \geq 10$ km and $7.5 \times 10^{-3}$~yr$^{-1}$ for $D_{\rm i} \geq 1$ km. Using $\alpha_2 = -2.1$ we find similar results: $9.3\times10^{-5}$~yr$^{-1}$ for $D_{\rm i} \geq 10$ km and $1.1\times10^{-2}$~yr$^{-1}$ for $D_{\rm i} \geq 1$ km. We note that Case A of \citet{Zahnle2003} represents a shallower size-frequency distribution of the impactors, with $\alpha_3=-1$ for $D_{\rm i}<1.5$~km, whereas their Case B reflects a steeper distribution closer to a collisionally evolved population deduced from craters on Triton ($\alpha_3=-2.5$ for $D_{\rm i}<1.5$~km). Most recent studies adopt Case A as a more accurate representation of heliocentric comet populations in the outer Solar System, so that we shall do the same. The shallow slope is also more closely aligned with the results of \citet{Singer2019}.\\

\citet{Zahnle2003} theoretically assessed the impact rate on Saturn by considering several factors: the proportion of comets delivered from beyond Neptune’s orbit to Jupiter’s, the impact probability based on the planets’ relative sizes to their Hill spheres, and the gravitationally enhanced cross-sections, which are proportional to each planet’s size and escape velocity. From this, they estimated the impact ratio of Jupiter to Saturn to be 1:0.42. This is close to the ratio of 1:0.47 derived from the average annual \"{O}pik impact probabilities for the three then-known Centaurs with Saturn, for which the impact rate is computed as $2 \times 10^{-8}$~yr$^{-1}$ for a 150 km diameter object \citep{Fernandez2002}. Together this corresponds to a Saturnian impact rate of $3.2 \times 10^{-3}$~yr$^{-1}$ for $D_{\rm i} \geq 1$ km, using the Case A size-frequency distribution. \\

With this information we compute that the current impact rate on Enceladus, scaled from Jupiter, using Case A is approximately 16.7~Gyr$^{-1}$ for objects with $D_{\rm i}>1$~km. We used $P_{\rm s}/P_{\rm P}=2.2\times10^{-6}$ for the Enceladus to Jupiter impact probability ratio, and that the impact rate on Jupiter is $5 \times 10^{-3}$ yr$^{-1}$ for $D_{\rm i} \geq  1.5$ km. This impact rate on Enceladus is a factor of eight higher than our estimate, and is primarily due to \citet{Zahnle2003} taking a factor of six higher impact rate on Jupiter, as well as our different size-frequency distribution between impactor diameters 1~km and 1.5~km. To compute the cratering rate on Enceladus we use the inverse of the crater scaling laws adopted by \citet{Zahnle2003} (see the Appendix). Using the impact velocity of heliocentric planetesimals with Enceladus listed by \citet{Zahnle2003} of $v_{\rm i}=24$~km~s$^{-1}$ the impactor diameter required to excavate a crater with diameter $D_{\rm cr}>10$~km on Enceladus is $D_{\rm i}=0.34$~km. Adopting the Case A size-frequency distribution implies the current impact rate of such planetesimals on Enceladus is 49~Gyr$^{-1}$ and the resulting cratering rate is then $6.1 \times 10^{-5}$~km$^{-2}$~Gyr$^{-1}$, which is a factor of 12 greater than our estimate, with the additional factor of two over the impact rate being caused by our different adopted crater scaling law and slope. \\

\subsection{Recent studies} 
\label{sub:recent_studies}
In the last five years, \citet{Nesvorny2023} have built upon their earlier works of \citet{Nesvorny2015a,Nesvorny2015b,Nesvorny2017b} to further constrain the initial conditions of the giant planet migration model, specifically the initial semi-major axis of Neptune and its migration pathway. The most recent version of the model assumes that the dynamical instability among the giant planets occurred 10~Myr after the dispersal of the protoplanetary gas disk. The migration rate of Neptune followed an exponential profile \citep{Nesvorny2017b}: for the first 10 Myr, the migration e-folding time ($\tau$) is 10 Myr, representing rapid migration. Then, after 10 Myr of dynamical evolution, $\tau$ increases to 30 Myr, reflecting a slower migration rate, which lasts up to 500 Myr.\\

\citet{Nesvorny2023} uses the initial conditions of the simulations of \citet{Nesvorny2017b} and reran the last 1~Gyr of evolution. They began with one million test particles presumably taken from a snapshot after 3.5~Gyr of evolution from \citet{Nesvorny2017b} and applied `on-the-fly' cloning when any particle ventured closer than 23~AU for the first time, cloning it 50 times to maintain sufficient particle interactions with the planets. \\

From the simulation's impact counts, \citet{Nesvorny2023} calculated the impact rate on Jupiter as $3.5 \times 10^{-5}$ per year for objects with $D_{\rm i} \geq 10$ km, without considering cometary disruption. They compute the impact rate as $\dot{C}_p = (\Delta N / T)~P_{\rm P}$, where $T$ is the duration of the simulation (1~Gyr), and $\Delta N$ is equal to the original number of planetesimals in the disc because 99.7\% are lost \citep{Nesvorny2017b}. \citet{Nesvorny2023} adopts $\Delta N = (8 \pm 3)\times10^9$ with $D_{\rm i}>10$~km. In the simulations they record 217 impacts on Jupiter, out of a total of 50 million test particles, implying $P_{\rm P}=4.34\times10^{-6}$. This value is three orders of magnitude lower than both our estimate and that of \citet{Levison2000a}, and the reason for this discrepancy is not immediately clear. Combined with the loss of planetesimals over the duration of the simulation the nominal value $\dot{C}_p =3.5 \times 10^{-5}$~per year is obtained. With cometary disruption included, the rate decreases to $2.6 \times 10^{-5}$~yr$^{-1}$. Using their adopted impactor power-law slope of $\alpha_2=-2.1$, we compute that the impact rate on Jupiter for $D_{\rm i} > 1.5$ km becomes $1.9 \times 10^{-3}$~yr$^{-1}$ without disruption, and $1.4 \times 10^{-3}$~yr$^{-1}$ with disruption. These values are still higher than ours by factors of a few, but they also fall outside the range of $5^{+6}_{-3} \times 10^{-3}$~yr$^{-1}$ cited from \citet{Zahnle2003}. The impact rates for objects with $D_{\rm i}>1$~km are listed in Table~\ref{tab:rates}. However, the high loss rate reported in the simulations of \citet{Nesvorny2023} is unexpected, as our simulations indicate a loss of 80\% of planetesimals over the last 3.5~Gyr after a dynamical instability 4.5~Gyr ago. \\

To calculate the collision probability with Enceladus, \citet{Nesvorny2023} recorded the close encounters within a Hill radius of Saturn and numerically evaluated the collision probability for each Saturnian encounter with Enceladus. Following \citet{Zahnle2003} the impact rate was then scaled to that of Jupiter. 
Using their nominal impact rate on Jupiter of $3.5\times10^{-5}$~yr$^{-1}$ for $D_{\rm i} \geq 10$ km and $P_{\rm s}/P_{\rm P}=1.5\times10^{-6}$ as the Enceladus to Jupiter impact probability ratio -- see Table~2 of \citet{Nesvorny2023} -- the impact rate on Enceladus becomes 0.053~Gyr$^{-1}$ for $D_{\rm i}>10$~km, or 6.6~Gyr$^{-1}$ for $D_{\rm i}>1$~km using $\alpha_2=-2.1$; this impact rate is a factor of 2.5 higher than our value. \\

\citet{Nesvorny2023} does not explicitly compute the cratering rate onto the satellites, only the relative impact probabilities. Since some of their methodology follows that of \citet{Zahnle2003} we can naively adopt the same cratering scaling law as \citet{Zahnle2003} and the size-frequency distribution of their Case A for objects with diameter $D_{\rm i}<1$~km. We then compute the cratering rate on Enceladus as $2.4 \times 10^{-5}$~km$^{-2}$~Gyr$^{-1}$ for the non-disruption case and $2.2 \times 10^{-5}$~km$^{-2}$~Gyr$^{-1}$ with disruption. These cratering rates are a factor of three to four higher than ours and a factor of three to four lower than that of \citet{Zahnle2003}.

\section{Discussion}
In this section we discuss our results in comparison to previous studies in more detail.\\

We find that the impact rates onto Jupiter and Enceladus, and by extension the other giant planets and satellites, is lower than that found by two other studies. In this work we have made our methodology transparent, and we clarified every choice of every parameter that we have used and how they are used. It is therefore strange that despite many of these parameters being similar to some of the previous studies, we still find a very different cratering rate. Why is this so? There are several reasons that we can think of. \\

The first is the outcome of the numerical simulations. The migration simulations in \citet{Wong2023} move Neptune and Uranus out slower than \citet{Nesvorny2017b}, but \citet{Nesvorny2023} concluded that the details of the migration should not matter for the current impact rates. We concur with this statement based on logical deduction alone. However \citet{Wong2023} found that the disc's width appears to play a role because it influences the depletion, fraction that becomes ecliptic comets and the rate of decay, but only when the outer edge is beyond 35~au. Furthermore, the migration occurred very early, and the outer solar system has been relatively stable for the past 3.5 Gyr. Therefore the current impact rate -- which reflects only the most recent few hundred Myr -- should not strongly depend on the specific details of Neptune's migration, such as its speed or timing if it occurred more than 4 Gyr ago. The adopted disc mass and its width in \citet{Wong2023} are also comparable to that of \citet{Nesvorny2017b}, and we obtain the same amount of depletion by 4.5~Gyr of evolution, apart from cases where the disc extended beyond 35~au. The average rate of decline of our scattered disc population is consistent with other works in the literature so that if this parameter was fundamentally wrong in our simulations it has to be equally wrong in others; this is highly unlikely. Unfortunately we do not have the numerical resolution to determine $f_{\rm JFC}$ and other studies do not report $f_{\rm EC}$ so that we cannot compare our values to the works of others; the only parameter with which we can do this is $r_{\rm SD}$, which agrees across the board. \\

One difference between the outcome of our numerical simulations and that of \citet{LevisonDuncan1997} and \citet{Levison2000a} is that our impact probability with the planets is lower than reported in these studies, even though the impact probability for the simulations with migration included are comparable to those in \citet{Levison2000a}. Even though \citet{LevisonDuncan1997} do not explicitly report the impact probabilities, the simulations of \citet{Levison2000a} also find a comparable lifetime for the ecliptic comets than us, once again lending credibility to our approach. We have also argued why our clone simulations are a good proxy for the current population and why the impact probabilities from this population should be considered rather than that including the migration. In our studies, we derive the impact probability probabilities of the satellites by multiplying the impact count during the flyby simulations with the encounter probability with the planets. Both counts were recorded in dynamic simulations. \citet{Nesvorny2023}, however, numerically evaluated the collision probability of each planetesimal encounter with the giant planets and its satellites, yet we find similar impact probability ratios $P_{\rm s}/P_{\rm P}$ to \citet{Zahnle2003} and \citet{Nesvorny2023} so that our different approach still yields consistent results.\\

A second potential reason for disagreement is the current number of scattered disc objects that we have assumed. Neither \citet{Zahnle2003} nor \citet{Nesvorny2023} explicitly mention this number because they have no need for it. Unfortunately this quantity is poorly known and it will probably remain so for the time being. We have argued why we have chosen the number that we did, based on constraints from the total depletion of the population of planetesimals in our simulations as well as constraints on the primordial disc mass and the observed slopes of the size-frequency distribution of these objects, and the observational constraints. We do not foresee that the current population can be much more or much less numerous than what we have used here; factors of two perhaps. As such, this quantity is perhaps also not to blame.\\

A third option is the choice of impactor size-frequency distribution. Our choices are certainly different from those of \citet{Zahnle2003} and partially accounts for our lower impact rates, but we deliberately chose the same slope $\alpha_2=-2.1$ as \citet{Nesvorny2023} for objects with diameter $1\lesssim D_{\rm i} \lesssim 100$~km to minimise differences in outcome. We have extended the population to sub-km objects by adopting $\alpha_3=-0.7$ from \citet{Singer2019}, but if we restrict ourselves to computing the impact rate of km-sized objects then this choice does not matter. Thus our factor of a few difference with \citet{Nesvorny2023} cannot be explained by the choice of impactor size-frequency distribution slope. \\

Beyond these relatively obvious potential differences that all yield comparable outcomes, there is one additional divergence in impact rate calculation methods that could contribute to the observed discrepancies in our results: the fundamental difference in approach. Both \citet{Zahnle2003} and \citet{Nesvorny2023} anchor their satellites' impact rate to that of Jupiter. Whether this is derived from simulations or from observational constraints is less relevant. The satellite-to-Jupiter impact probability ratio and Jupiter's absolute impact rate yield the resulting satellites' impact rate. While the impact probability with satellites is implicit in this method, it is indirectly informed by the impact probability with Jupiter, even if this is not explicitly stated in \citet{Zahnle2003}.\\

In contrast, our approach uses the flux of new comets entering the Centaur region from beyond Neptune as the fundamental parameter. This is then multiplied by the absolute impact probability with the satellites -- derived from flyby simulations and the {\it encounter} probabilities with the giant planets -- to yield the satellites' impact rate. This is where the major differences show up: our impact rate so computed on Jupiter is a factor of eight lower than \citet{Zahnle2003} and four lower than \citet{Nesvorny2023}. The discrepancy with \citet{Zahnle2003} is primarily due to their choice of fixing the impact rate on Jupiter from historical observations rather than numerical simulations, while the discrepancy with \citet{Nesvorny2023} is more difficult to explain, but it appears to be due to how they compute the impact rate on Jupiter over the last billion years for objects with $D_{\rm i}>10$~km, and their potentially anomalously low impact probability with this planet.\\

As a thought experiment we may use the impact rate on Jupiter adopted by \citet{Zahnle2003} at face value and compute what the injection rate of comets with $D_{\rm i}>1$~km from beyond Neptune needs to be to be consistent with this impact rate. We obtain $F_{\rm EC}=2.6$~yr$^{-1}$. For this high value there is no literature precedent. Even if we were to adopt an impact probability with Jupiter of about 1\% \citep{Levison2000a} rather than our value of 0.29\%, the injection rate still needs to be 0.75~yr$^{-1}$, which is still very high. Repeating this check for the impact rate of \citet{Nesvorny2023} with our impact probability and without disruption results in 1.5~yr$^{-1}$, for which there is also no literature precedent either; using the higher impact probability on Jupiter of 1\% lowers this to 0.44~yr$^{-1}$, which is acceptable, but this cannot be verified without knowing the impact probaility with Jupiter.\\

One potential effect that we did not consider is tidal disruption of comets. From crater chains on Ganymede and Callisto and assuming their surfaces ages are about 4~Gyr old \citet{Schenk1996} argue that there are $3.7\times10^{-3}$ tidal disruptions per year near Jupiter. Since these disruptions generally happen within 3~$R_J$ for comets with $D_{\rm i}\sim 1$~km this rate of disruption is fairly consistent with our estimated impact rate on Jupiter, yet with typically 10 fragments generated these disruptions are insufficient to elevate the impact rate at Jupiter by a factor of a few.\\

\section{Conclusions}
In this work, we used numerical simulations of the evolution of the outer solar system to compute the impact and cratering rates onto the regular satellites of Jupiter, Saturn, and Uranus, highlighting the complexities involved in determining these rates. We demonstrated how the current cratering rate depends strongly on the specific parameters of the dynamical/disc model employed.\\

We outlined our methodology and parameter choices in detail, allowing future studies to verify or adjust these assumptions and recalculate impact rates. After careful comparison with previous works, we conclude that our impact rates are lower than that the best estimates of previous studies by up to a factor of 6, likely due to our reliance on the injection rate of scattered disc objects into the Centaur region rather than directly anchoring to the impact rate on Jupiter. This highlights the importance of clearly stating all parameters and assumptions in future studies to enable meaningful comparisons.\\

Our findings underscore that the uncertainties in impact rates significantly influence surface age estimates, particularly for young, sparsely cratered terrains such as Europa, Enceladus, and young patches on Tethys and Dione. Better constraints on orbital dynamics and impactor size-frequency distributions are needed to refine both current and historical cratering rates.\\

Future advances, such as more precise measurements of small body sizes using stellar/solar occultation techniques, can provide crucial data to constrain the current impact rates and improve our understanding of the historical evolution of small bodies in the outer solar system. These insights will help bridge gaps between impact dynamics and geological interpretations of icy moon surfaces.

\section*{Acknowledgments}
This study is supported by the Research Council of Norway through its Centres of Excellence funding scheme, project No. 332523 PHAB. GENGA can be obtained from \url{https://bitbucket.org/sigrimm/genga/}
\bibliographystyle{aa}
\bibliography{main}
\begin{appendix}
\section{Crater scaling}
\citet{Zahnle2003} uses the following crater scaling law

\begin{eqnarray*}
    D_{\rm cr}&=&4.38\left(\frac{D_{\rm i}}{1\,{\rm km}}\right)^{0.783} \left(\frac{\rho_i}{\rho_s}\right)^{0.333}\left(\frac{v_{\rm imp}}{1\,{\rm km\,s}^{-1}}\right)^{0.434}\left(\frac{g}{1\,{\rm m\,s}^{-2}}\right)^{-0.217} \nonumber \\
    D_{\rm cr}&=&5.31\left(\frac{D_{\rm i}}{1\,{\rm km}}\right)^{0.885} \left(\frac{\rho_i}{\rho_s}\right)^{0.376}\left(\frac{v_{\rm imp}}{1\,{\rm km\,s}^{-1}}\right)^{0.490}\left(\frac{g}{1\,{\rm m\,s}^{-2}}\right)^{-0.245} \nonumber \\
    &\times&\left(\frac{D_{\rm SC}}{1\,{\rm km}}\right)^{-0.13}
\end{eqnarray*}
where $g$ is the acceleration due to gravity and $\rho_i$ and $\rho_t$ are the densities of the impactor and the satellite, and $D_{\rm SC}$ is the simple to complex transition crater diameter. \citet{Zahnle2003} uses $D_{\rm SC}= 2.5$~km for Europa, Ganymede, Callisto, and Titan \citep{Schenk2004} and $D_{\rm SC}= 15$~km for all other satellites as well as an impactor density of $\rho_{\rm i}= 600$~kg~m$^{-3}$ and a satellite density of $\rho_{\rm s}= 900$~kg~m$^{-3}$ for all satellites apart from Io, for which $\rho_{\rm s}= 2700$~kg~m$^{-3}$. The top equation should be applied when $D_{\rm cr}<D_{\rm SC}$. The inverse equations are then

\begin{eqnarray*}
    D_{\rm i}&=&0.151\left(\frac{D_{\rm cr}}{1\,{\rm km}}\right)^{1.28} \left(\frac{\rho_i}{\rho_s}\right)^{-0.425}\left(\frac{v_{\rm imp}}{1\,{\rm km\,s}^{-1}}\right)^{-0.554}\left(\frac{g}{1\,{\rm m\,s}^{-2}}\right)^{0.277} \nonumber \\
    D_{\rm i}&=&0.151\left(\frac{D_{\rm cr}}{1\,{\rm km}}\right)^{1.13} \left(\frac{\rho_i}{\rho_s}\right)^{-0.425}\left(\frac{v_{\rm imp}}{1\,{\rm km\,s}^{-1}}\right)^{-0.554}\left(\frac{g}{1\,{\rm m\,s}^{-2}}\right)^{0.277} \nonumber \\
    &\times&\left(\frac{D_{\rm SC}}{1\,{\rm km}}\right)^{0.147}
\end{eqnarray*}

\end{appendix}

\end{document}